\documentclass{aa}
\usepackage{epsf}
\begin{document}

%
\titlerunning{Statistical properties of exoplanets IV}
\title{Statistical properties of exoplanets IV. The period--eccentricity
relations of exoplanets and of binary stars}
\author{J.L.~Halbwachs\inst{1}
   \and M.~Mayor\inst{2}
   \and S.~Udry\inst{2}}

\offprints{J.L. Halbwachs}
\institute{
   Observatoire Astronomique de Strasbourg (UMR 7550),
   11 rue de l'Universit\'{e}, F--67\,000 Strasbourg, France \\
   email: halbwachs@astro.u-strasbg.fr
 \and Geneva Observatory,
   51 chemin des Maillettes, CH-1290 Sauverny, Switzerland \\
   email: Michel.Mayor@obs.unige.ch, Stephane.Udry@obs.unige.ch
}
\date{Received 3 May 2004; accepted 19 October 2004}
\abstract{A sample of spectroscopic binaries and a sample of single planetary
systems, both
having main-sequence solar-type primary components, are selected in order
to compare their eccentricities. The positions of the objects in the
($P.(1-e^2)^{3/2}$, $e$) plane is used to determine parts in the
period--eccentricity diagram
that are not affected by tidal circularization. The
original eccentricities of binaries and planets are derived and compared.
They seem to be weakly or not at all correlated with period in both
samples, but two major differences are found :

\noindent
(1) The tidal circularization of planetary orbits is almost complete for periods
shorter than 5 days, but it is not visible when $P.(1-e^2)^{3/2}$
is longer than this limit. This suggests that the circularization
occurs rapidly after the
end of the migration process and is probably simultaneous with the end
of the formation of the planet. By contrast, we confirm
that the circularization of the
binary orbits is a process still progressing a long time after the formation
of the systems.

\noindent
(2) Beyond the circularization limit, the eccentricities of the orbits of the
planets are significantly smaller than those of binary orbits, and
this discrepancy cannot be due to a selection effect. Moreover, the
eccentricities
of binaries with small mass ratios are quite similar to those of all binaries
with $q<0.8$. This
suggests that the low eccentricities of exoplanet orbits
are not a consequence of low-mass secondaries in a universal process.

These remarks are in favor of
the idea that binaries and exoplanets are two different
classes of object from the point of view of their formation.

}
\maketitle
\section{Introduction}

It is well known that
the orbits of the exoplanets with periods larger than 5 or 6 days have
eccentricities significantly larger than those of the giant planets of the solar
system. Several mechanisms were proposed to explain this feature, but, up to
now, none is fully convincing. It was proposed that eccentric orbits could be a
consequence of the
dynamic evolution of systems initially involving several planets (Rasio \&
Ford \cite{RaFo96}, Lin \& Ida \cite{LinI97}, Ford et al. \cite{Ford01}, 
Papaloizou \& Terquem \cite{PapaTer01}, Rice et al. \cite{Rice03}), but
these models fail to produce the frequency of giant planets
with semi-major axes smaller than about 1 AU. The giant planets close to
their harboring stars are often assumed to be produced by migration
within a disk (Ward \cite{Ward97}, Masset \& Papaloizou \cite{MaPa03} and
references therein), but this process hardly produces
eccentric orbits (Papaloizou et al. \cite{PaNeMa01}, Thommes \& Lissauer
\cite{ThoLi03}), although Goldreich \& Sari (\cite{GolRei03}) and Woolfson
(\cite{Woolf03}) leave some room for hope. Therefore, it is
tempting to consider that the exoplanets
are generated by the same process as binary stars (Stepinski \& Black
\cite{SteBla00}). This implies that giant exoplanets were not formed
by gas accretion onto a heavy rocky core, as usually assumed, but by an
alternative process. They could come from disk instabilities
(Mayer et al. \cite{Mayer02}, Boss \cite{Boss02}, \cite{Boss03}), but
inward migration in a disk is then invoked again to explain the short-period
orbits; alternatively, planets could be generated by fragmentation
of a collapsing protostellar cloud, {\it via} filament condensation and
capture (Oxley \& Woolfson \cite{OxWoo04}), or even exactly as stellar
components in binary systems
(see discussion in Bodenheimer et al. \cite{BoHuLi} and references therein).
However, these models may be efficient in forming massive planets or brown
dwarfs, but
not planets around 1 Jupiter mass or less.

Note that the binary formation models are not very
satisfactory either (see the review by Tohline \cite{Toh02}).
The large eccentricities of binaries are explained
by fragmentation of collapsing cores and subsequent interactions
between the forming stars
(Bate et al. \cite{BaBoBro02}, Goodwin et al. \cite{GoWhiWa04}), but, as
for exoplanets, the simulations do not provide the high frequency of close
systems. 
Moreover, statistical investigations on main sequence binaries 
(Halbwachs et al. \cite{Halb03}, Paper I hereafter)
have shown that the close binaries (i.e. with semi-major axes less than
a few AU) consist in two populations~: one
with large eccentricities and mass ratios less than 0.8 (``non-twins''
hereafter), and one with
moderate eccentricities and nearly identical components (``twins'').
Additionally, the
twins are more frequent among short-period binaries than among the others.
At first, these properties were derived from binaries with F7--K primary
components, but they are also valid for M-type dwarfs (Marchal et al.
\cite{MaDel03}).

In the present paper, the period--eccentricity diagram is used to compare
the exoplanets with the binary stars~: our main purpose is to investigate if
the properties of exoplanets may be considered as an extrapolation of the
properties of binaries in the range of very low mass ratios. This would
indicate whether the formation processes of these objects are similar.
In the course of the paper, a few other points are also treated~: (1)
the correlation between the eccentricity and the period or the angular momentum,
(2) the relation between the eccentricity and the
metallicity of planets, (3) the original distributions of eccentricities
for planets and for binaries, considering the twins separately.
Comparisons between planets and binaries in the period--eccentricity diagram
were already presented by Mayor et al. (\cite{Mayor01}), Mazeh \& Zucker
(\cite{MaZu01}), and Udry (\cite{Udry01}), who concluded that planets and
binaries are very similar when periods longer than 50 days are considered.
However, their samples contained around 30 or 40 planets, and a many others
have been discovered since. The question needs to be re-considered.

The interpretation of the period--eccentricity diagram is rather complex,
and our investigations are based on the method presented in Sect.~\ref{method}. 
Sect.~\ref{binaries} is dedicated to the binaries; we investigate if,
additionally to the twins, other classes of mass ratio have specific
distributions of eccentricities.
A similar treatment is applied to
exoplanets in Sect.~\ref{planets}. Binaries and exoplanets are compared in
Sect.~\ref{comparison}, in which we
pay attention to the difference in the selection effects of both samples.
The consequences of our results are discussed in Sect.~\ref{conclusion}.

\section{Method}
\label{method}

\subsection{Tidal effects}
\label{tides}

We must pay attention to the fact that the periods ($P$) and the
eccentricities ($e$) of the objects are modified by tidal interactions,
especially when $P$ is short.
As a consequence, the ($P$ -- $e$) diagram may schematically be divided into two
parts~: the short periods, where the orbits are circular or have low
eccentricities, and the periods 
longer than the circularization limit, hereafter called $P_{\rm cutoff}$
(Mayor \& Mermilliod \cite{MaMer84}, Duquennoy \& Mayor \cite{DM91}).
Several theoretical models were proposed to derived $P_{\rm cutoff}$, and the
treatment is not the same for
binaries (Zahn \cite{Zahn92}, Hut \cite{Hut81}, \cite{Hut82},
Keppens \cite{Keppens97}, and references therein)
and for planets
(Goldreich \& Soter \cite{GoSo66}, Trilling \cite{Trilling00}). Moreover,
several physical processes are invoked, each of them depending differently
on the mass ratios of the systems.

\begin{figure}
\epsfxsize=8. cm \epsfbox[25 95 455 410]{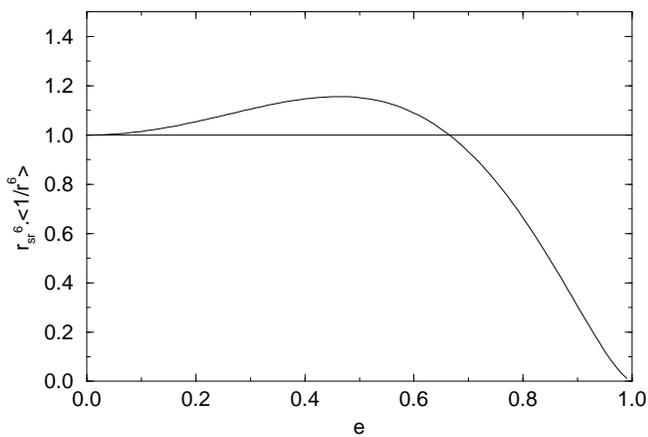}
  \caption[]{The mean value of the $1/r^6$ term of the tidal torque,
compared to
$1/r_{\rm sr}^6$ , where $r_{\rm sr}$ is
the semilatus rectum.  Except for very large
eccentricities,
$1/r_{\rm sr}^6$ is
an acceptable approximation of $<1/r^6>$.
}
  \label{r6}
\end{figure}

Despite the complexity of the process, a few simple guidelines may be 
drawn. First of all, the
efficiency of tides in modifying the orbits is very sensitive to the
distance between the components.
For a given system, the tidal torque depends on the orientation of the tidal
bulge and on the separation between the components, $r$. It varies as
$1/r^6$ (Lecar et al. \cite{LeWhe76}).
Therefore, for systems differing only by
period, the transition from circularized orbits to orbits
practically unaffected by tides is a narrow strip in the ($P$ -- $e$)
diagram (see the simulations by Witte \& Savonije, \cite{WiSa02}).
However, this does not mean that the systems with $P>P_{\rm cutoff}$ may have
any eccentricity. For a given period the systems with eccentric orbits
have components much closer than the semi-major axis during a part of
the period, and below a
certain limit the orbit rapidly becomes circular. As a consequence,
the upper part of the ($P$ -- $e$) diagram is cleared even for
$P>P_{\rm cutoff}$. Note that the orbits that were originally eccentric
do not keep the same period when they are evolving towards $e=0$~:
when the primary star is a slow rotator, the orbit is circularized keeping
the orbital angular momentum unchanged (Witte \& Savonije \cite{WiSa02}, Hurley
et al.  \cite{HurToPo02}). Therefore, the semilatus rectum, $r_{\rm sr}=a(1-e^2)$,
is conserved and it becomes the radius of the final
circular orbit. As a consequence, the final period is~:

\begin{equation}
P_{\rm sr} = P.(1-e^2)^{3/2}
\label{Pcirc}
\end{equation}

where $P$ and $e$ refer to the original state of the system. This gives us 
a simple but efficient way to explore the transition from circularized
orbits to orbits unaffected by tidal effects. In place of a
($P$ -- $e$) diagram, the objects are plotted in the ($P_{\rm sr}$ -- $e$)
plane. Therefore, the evolution path toward a circular orbit is a vertical
line in the diagram. Moreover, note that, for a wide range of
eccentricities, the mean value of $1/r^6$ for a complete orbit is approximately
$1/{{r_{\rm sr}}^6}$ (Fig.~\ref{r6}). As a
consequence, we expect that, when $r_{\rm sr}$ is small enough to permit efficient
tidal effects for a given eccentricity, these effects will remain important
during the evolution of the orbit, until they eventually lead to
circularization.
Therefore, the border between the circularized systems and the
area nearly unaffected by tides in the ($P_{\rm sr}$ -- $e$) diagram should appear
very clearly. If the circularization were not at all related to the ages of
the systems, the diagram should show a strong contrast, with the circular orbits
on the left--hand side, and the orbits with any eccentricities, including
the largest ones, immediately beyond the circularization limit. This is almost
what is observed in reality, especially for the exoplanets
(see Fig.~\ref{Psr-e_SB} and~\ref{Psr-e_pla} below).

In practice, however, a real sample is selected up to a maximum period
$P_{\rm Max}$, and not all the eccentricities are permitted in the
($P_{\rm sr}$ -- $e$)
diagram. For a given $P_{\rm sr}$, the eccentricities range from 0 to the
limit~:

\begin{equation}
e_{\rm Sup} = \sqrt{1 - (P_{\rm sr}/P_{\rm Max})^{2/3}}.
\label{eSup}
\end{equation}

With $P_{\rm cutoff}$ so determined, we derive from Eq.~(\ref{Pcirc})
the maximum eccentricity that
the systems unaffected by tides in the ($P$ -- $e$) diagram may have~: 

\begin{equation}
e_{\rm Max} = \sqrt{1 - (P_{\rm cutoff}/P)^{2/3}}.
\label{eMax}
\end{equation}

For comparing two samples which were have been differently affected by tidal
circularization, it is necessary to restrict the comparison to the
smallest of the two limits in $e_{\rm Max}$ in order to find discrepancies
coming only from the original distributions of $e$.

\subsection{Comparison of two samples}

The comparison between different samples of orbiting
systems is based on the median eccentricities. Two approaches are used~:
the first is a visual examination, and the second is a statistical test.
In the first approach,
the median eccentricity of each sample is derived as a function of
the period. In practice, for any period $P_i > P_{\rm cutoff}$,
the median eccentricity is derived from 6 systems taken among the periods
closest to $P_i$. Except near the limit of
the period range, 3 of these 6 systems have $P<P_i$ and the 3 others have
$P>P_i$. The second approach
is the statistical test used in Paper I~: the two samples are merged, and the
range of periods longer than $P_{\rm cutoff}$
is divided into bins, each containing 12 systems, except
for the last one which may contain up to 23 systems. The common median
eccentricity is derived in each bin, and the systems below the median are
counted for one of the two samples. If this sample actually belongs to the same
statistical population as the other, the probability of getting any
count, $P(k)$, obeys the hypergeometric distribution. The rejection
threshold of the null hypothesis, $H_0$~: ``all the systems are equivalent from
the point of view of the eccentricities'' is then derived. When the count $k$ is
less than half the population of the considered sample, the rejection
threshold of $H_0$ in a two-sided test is $2 \times P(i\le k)$;
on the contrary, it is $2 \times P(i\ge k)$ when k is larger than the expected
number.

Note that setting the content of the bins to 12 systems is a bit
arbitrary. This number is
neither too small to derive a reliable median nor too large to
have a nearly constant period distribution within each bin.
Using other numbers close to 12 would give other results, but
it was verified that the differences are not important.

\section{The binaries}
\label{binaries}

\subsection{The binary sample}

The so--called {\it extended sample} of F7--K dwarf binaries selected 
in Paper~I is used again. It consists of 89
spectroscopic binaries (SB) found in the solar neighborhood or in open
clusters, with periods of up to 10 years.
We already know that the twins have, on average,
eccentricities smaller than the other binaries. However, before comparing
binaries to exoplanets, it is worthwhile to see if the eccentricities of
non-twin binaries depend on the mass ratios. 

The mass ratios $q= {\cal M}_2/
{\cal M}_1$ of the SB in the sample have been fixed for 58 binaries, thanks to
the combination of the SB orbital elements with Hipparcos astrometric
observations, or with photometric sequences in the open clusters (Paper I).
For the other 31 SB, we derive intervals containing the actual mass
ratios. The minimum limits are computed from the mass functions; 
the maxima are obtained differently for the nearby SB and for the
cluster SB~: the former all have $q < 0.65$, since otherwise they would
be double--lined SB with known $q$, and the latter have limits
coming from their positions in the photometric sequence of the cluster. 

In order to make visible a possible relation between the mass ratios
and the eccentricities, the SB are distributed in several groups~:
$q \le 0.40$ (16 SB), $0.40 < q \le 0.80$ (20 SB), 
and twins (27 SB); we still add a
group containing all the SB with $q \le 0.80$ (62 SB, including the
36 already in the first two groups).

\subsection{Limit of tidal circularization}

\begin{figure}
\epsfxsize=8. cm \epsfbox[25 95 455 410]{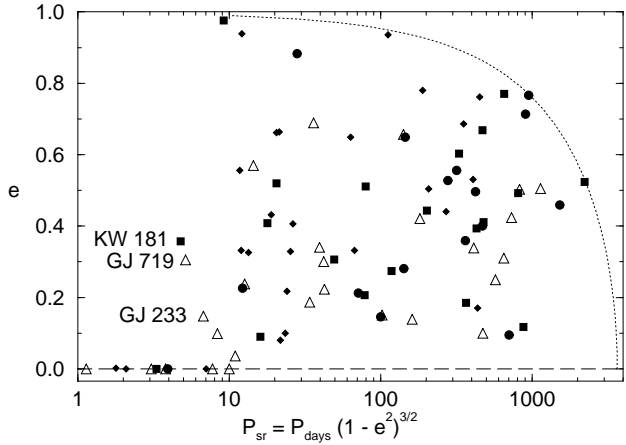}
  \caption[]{Distribution of the SB in
the ($P_{\rm sr}$ -- $e$) diagram, where $P_{\rm sr}$ is the period that the
SB would have if their orbits became circular with the same angular
momenta. The symbols represent the systems with different mass ratios.
The code is as follows~:
$q \leq 0.4$, circles; $0.4 < q \leq 0.8$, squares;
$q \leq 0.8$ (maximum $q$ for SB1) and minimum $q \leq 0.4$,
small diamonds;
$q > 0.8$ (twins), open triangles. 
The thin dotted line is the limit $e_{\rm Sup}$
corresponding to a maximum period of 10 years, as derived from Eq.~(\ref{eSup}).}
  \label{Psr-e_SB}
\end{figure}

The SB are
plotted in the ($P_{\rm sr}$ -- $e$) diagram (Fig.~\ref{Psr-e_SB}), to
delimit the part of the diagram affected by tidal circularization. 

Several relevant features appear in Fig.~\ref{Psr-e_SB}. First of all, the range
of eccentricities jumps from 0 to almost 1 between 4 and 9 days. All SB with
$P_{\rm sr}$ shorter than 4 days are at present on circular orbits, and
no evidence of circularization is visible when $P_{\rm sr}$ exceeds 10 days.
The SB with $P_{\rm sr}$ between 4 and around 10 days have medium
eccentricities, and some of them are even circularized.

At least one of the short period SB may have been generated with a very large
eccentricity, and may be too young for to have been circularized~:
\object{KW 181} has the most eccentric orbit among the periods shorter
than 10 days. This system belongs to the Praesepe cluster, and its age is
therefore only 0.8 $10^9$ years. Its circularization will be complete
within only 0.1 $10^9$ years (Duquennoy et al. \cite{DMM92}).
The other two SB with $e>0.1$ and $P>10$~days
may have eccentricities due to
perturbations by a third component (Kozai mechanism or secular perturbations,
see Mazeh \& Shaham \cite{MaSha79})~:
\object{GJ 719} has a CPM companion with a minimum separation of 280 AU
(Zuckerman et al. \cite{Zucker97}), and
\object{GJ 233} is a visual binary with a possible period of 200 years
(Heintz \cite{Heintz88}). When these 3 SB are discarded, we find 
$P_{\rm cutoff} \approx 8$ or 10 days, as in Duquennoy \& Mayor (\cite{DM91}).

Nevertheless, it is striking that
all the SB with eccentric orbits and $P<10$~days have large mass ratios.
In order to see if this feature is significant we look at the
period--eccentricity diagram of the 205 SB found by Latham et al.
(\cite{Latham02}) among stars with large proper motions. They also found
a range of period where circular orbits and eccentric orbits both
exist, but between around 10 and 20 days (see their Fig.~9). It is visible
in their plot that the double-lined SB (SB2), which have the largest $q$,
are not abnormally frequent
among the systems with
eccentric orbits and short periods. Therefore, we admit that the frequency of
twins with large eccentricity and $P<10$~days is just due to chance.

The sample of Latham et al. contains a few SB with circular orbits and periods 
between 10 and 20 days. However, they are stars with large proper motions, and
they generally belong to the old galactic disk or even to the halo. Therefore,
the long periods of some circularized orbits may be an effect of the ages of the
systems, in agreement with theoretical predictions (see e.g. Duquennoy et al.
\cite{DMM92}).
Moreover, these old stars are not representative of the stars
observed for planet detection.
For the parent population of stars
harboring planets, the limit of the area affected by
tidal effects in the ($P_{\rm sr}$ -- $e$) diagram is $P_{\rm cutoff}=10$~days.

\subsection{The binaries in the ($P$ -- $e$) diagram}

\begin{figure}
\epsfxsize=8. cm \epsfbox[25 95 455 410]{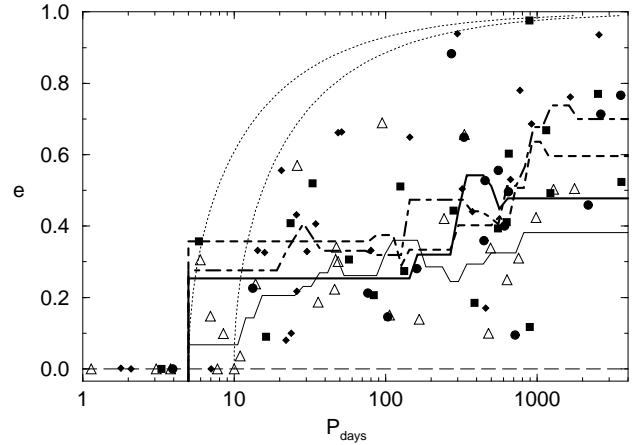}
  \caption[]{Distribution of the SB in
the period--eccentricity diagram. The symbols are the same as in
Fig.~\ref{Psr-e_SB}. The lines refer to the median eccentricities of the
different classes of mass ratios~:
$q \leq 0.4$, thick full line; $0.4 < q \leq 0.8$,
thick dashes; all $q \leq 0.8$, thick dot--dashed line;
$q > 0.8$ (twins), thin line. The 2 thin dotted lines are the
limits of the location of SB with $P_{\rm sr}$ between 5 and 10 days,
as derived from Eq.~(\ref{eMax}).
}
  \label{figP-eSB}
\end{figure}

The ($P$ -- $e$) diagram of the SB is plotted in Fig.~\ref{figP-eSB}. The
values of the median of the four classes of $q$ are drawn in this figure for
visual comparison.

As shown in Paper I, the twins often have below-average eccentricities
when periods longer than 5 days are considered;
the probability to get so large a discrepancy by chance, as derived by
the two-sided test, is only 2.7~\%.
By contrast, the
two groups with $q \le 0.80$ have nearly the same
distribution of eccentricities~: the significance of the two-sided
test is 100~\%. In order to check that even the SB with the smallest
mass ratios have the same eccentricity distribution as the others, the test
is done again by comparing the SB with $q \le 0.25$ (10 SB with $P > 10$ days)
to those with $0.25 < q \le 0.80$ (32 SB with $P > 10$ days). Again, exactly
half of the
low-mass ratio SB are below the median, providing a 100~\% significance.
Therefore, all the non-twin SB
may be considered together in the comparison to the exoplanets.

We now come back to the distribution of $e$ of all SB.
It is clearly visible in Fig.~\ref{figP-eSB} that the median eccentricity
increases with the period. However, we suspect that this may be
entirely explained by tidal circularization, because the orbits having
initially $P > P_{\rm cutoff}$ and $e > e_{\rm Max}$
are now circular with $P = P_{\rm sr} < P_{\rm cutoff}$.
In other words, we want to see if
the distribution of eccentricities depends on the periods,
apart from the cut--off at $e=e_{\rm Max}$.
A Spearman test is performed to check this hypothesis.
In order to discard the area affected by tidal circularization,
the test is restricted to the rectangular box ($P > 20\: {\rm d}, e < 0.61$).
The non-twins and the twins are
considered separately. The Spearman correlation coefficient is
0.26 for the former, and 0.29 for the latter;
taking into account the numbers of objects,
the probabilities to get by chance values so far from zero are
10~\% and 20~\% respectively.
Although these levels of significance are a bit low, they are still too large to
reject the hypothesis that the eccentricities are not correlated with the
periods as soon as the semilatus rectum of the orbit is larger than the
radius of a circular orbit with period $P_{\rm cutoff}$.

\subsection{Correlation eccentricity versus angular momentum}

Since $P_{\rm sr}$ is related to the angular momentum of the orbit, it seems
relevant to see if the eccentricities are correlated with this parameter.
This question looks similar to the ($P$ -- $e$) correlation investigated
just above, but it is different in reality, since changing the ($P$ -- $e$)
plane into the ($P_{\rm sr}$ -- $e$) plane modifies the density in relation with
the distribution of the periods. Therefore, the absence of correlation between
$P$ and $e$ does not necessarily imply an absence of correlation between
$P_{\rm sr}$ and $e$, and vice versa.

A Spearman test is used again, but in a box in the ($P_{\rm sr}$ -- $e$)
diagram. Since the selection of the sample was limited by the condition
$P<10$~years, we must take into account the limit $e_{\rm Sup}$ derived from
Eq.~(\ref{eSup}) (the thin dotted line in Fig.~\ref{Psr-e_SB}). Therefore,
the limits of the box considered in the Spearman test are 
$P_{\rm sr} \in [10\: {\rm d}, 789\: {\rm d}]$ and $e<0.8$, in order
to avoid the area with $e>e_{\rm Sup}$. Again, the rejection threshold
of the hypothesis that $e$ and $P_{\rm sr}$ are correlated is between 10 and
20~\%, and the eccentricities may not depend on the angular momenta of the SB.

\section{The exoplanets}
\label{planets}

\subsection{The planet sample}
We start from the up-to-date list of exoplanets that is provided on the Geneva
web site\footnote{http://obswww.unige.ch/Exoplanets}. 
However, several planets cannot be used in a comparison to the SB, for
several reasons~:
\begin{itemize}
\item
Since the binaries all have main-sequence primary components, the planets
orbiting subgiant or giant stars are discarded. The specific problem
of tidal circularization in a system containing an evolved star is thus avoided,
as well as the uncertainty coming from the evaluation of the mass of an evolved
primary component.
\item
The eccentricities must be reliable, and therefore derived from an orbit with
good quality. For that purpose, the residuals of the radial velocity (RV)
measurements, $rms$, are compared to the semi-amplitudes of the spectroscopic
orbit, $K$. Five planets having 
$rms$ larger than $K/3$ are discarded. This criterion looks a bit rough,
since several other factors could also be taken into account, such as
the number of measurements, the phase distribution, and 
the fact that some orbits refer to a second companion. Nevertheless, it
has the advantage of being simple, and it is worth noticing that all the SB
in our sample satisfy this condition.

Moreover, the planets that were not
followed by RV observations during a complete period are also discarded.
Since this last condition results in removing the majority of the planets with
periods longer than 2200 days, this value is adopted as a selection limit of
the sample.
\item
It seems that the planets found in binary or multiple stellar systems
have eccentricities smaller than the others when their periods are
less than 30 days (Eggenberger et al. \cite{EgUdMa04}). Three more planets are
removed from the sample for this reason.
\item
Multiple planetary systems are supposed to be different from those with
single planets, since the eccentricities may be affected by resonant
perturbations. It is worth noticing that the distribution of these
systems around the median eccentricity is not significantly different
from that of the single planets~: we find an excess of only one planet
with a low eccentricity (when an excess of large $e$ is expected), and the
probability to get this excess or a larger
one just by chance
is as large as 38~\%. Nevertheless, for security, we still discard 14 planets
belonging to multiple planetary systems.
\end{itemize}

A sample of
72 exoplanets orbiting main-sequence stars with periods
shorter than 2200 days and with reliable orbits is thus finally obtained.

\subsection{The exoplanets in the ($P_{\rm sr}$ -- $e$) diagram}

\begin{figure}
\epsfxsize=8. cm \epsfbox[25 95 455 410]{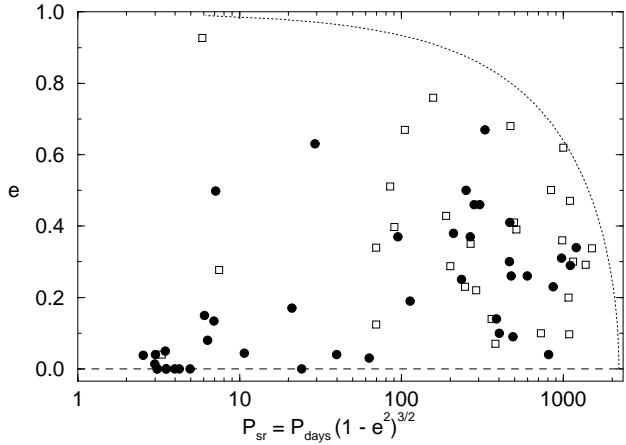}
  \caption[]{Same as Fig.~\ref{Psr-e_SB} for the planets.
The circles are the planets with minimum mass $< 2$~Jupiter, and the
open squares refer to the others. 
The thin dotted line is the
limit $e_{\rm Sup}$ corresponding to a maximum period of 2200 days.}
  \label{Psr-e_pla}
\end{figure}

As for the SB, the planets are plotted on the ($P_{\rm sr}$ -- $e$) diagram
(Fig.~\ref{Psr-e_pla}), in
order to investigate the effects of tidal circularization. The sample is
split into two nearly equal groups, one containing the planets with minimum
mass less than 2 Jupiter, and one with the planets heavier than this limit.
In contrast to the SB, for which circular orbits and moderate eccentricities
are mixed in a small range of $P_{\rm sr}$,
the separation between the circularized orbits and the others
is remarkably well determined, at $P_{\rm cutoff}=5$~days.
The most eccentric planetary orbit, \object{HD 80606b} (Naef et al. 
\cite{Naef01}), is actually found for this
period, but Wu \& Murray (\cite{WuMu}) demonstrated that it may be
excited by a distant companion through the
Kozai mechanism.

It appears in Fig.~\ref{Psr-e_pla} that the clustering of planets with
$P<10$~days (Udry et al. \cite{UdMaSan03}) is even more marked when $P_{\rm sr}$
is used in place of $P$~:
the planets are concentrated in orbits with the
semilatus recta corresponding to $P_{\rm sr}$ between 2.5 and 10 days, since we
count 18 planets in this range, but only 1 between 10 and 20 days.

It would be relevant to check if the eccentricities are correlated with
$P_{\rm sr}$, but our sample does not permit this~: the detection of the
planets is far from complete, and the incompleteness increases with the
period. Therefore, since the eccentric orbits correspond to a longer
period for a fixed $P_{\rm sr}$, the ($P_{\rm sr}$ -- $e$) diagram of the
planets is biased against large eccentricities. For that reason, the
($P_{\rm sr}$ -- $e$) diagram of the planets can be used only for investigating
the circularization limit. As a consequence, it cannot be used to compare
the SB to the exoplanets in Sect.~\ref{comparison}.

\subsection{The exoplanets in the ($P$ -- $e$) diagram}

\begin{figure}
\epsfxsize=8. cm \epsfbox[25 95 455 410]{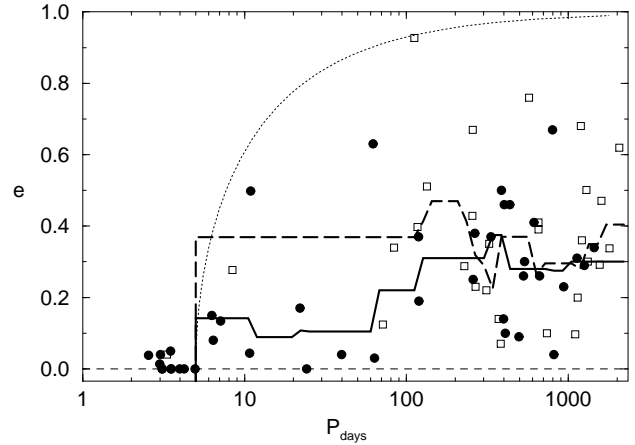}
  \caption[]{Same as Fig.~\ref{figP-eSB} for the exoplanets.
The symbols are the same as in Fig.~\ref{Psr-e_pla}. The full line
is the median eccentricity of the planets with
minimum mass $< 2$~Jupiter, and the dashes refer to the masses heavier
than 2 Jupiter. The thin dotted line is the
eccentricity $e_{\rm Max}$ corresponding to a circularized period of 5 days.}
  \label{figP-epla}
\end{figure}

The ($P$ -- $e$) diagram of the exoplanets is given in Fig.~\ref{figP-epla}.
Only 1 planet above 2 Jupiter masses has $P$ between 5 and 70 days.
This paucity of heavy-mass planets with
short periods has already been pointed out (Zucker \& Mazeh \cite{ZuMa02},
Udry et al. \cite{UdMaSan03}), and it makes the median eccentricity
of these planets unreliable in this range of period.
A two-sided test based on the common median for $P$ in the range 5 to 2200 days
shows that the probability
to get differences at least as large as that obtained is 60~\%.
It is then quite possible
that the eccentricities of planetary orbits are not related to the
masses of the planets.
This question is considered again in Sect.~\ref{secfe}.

The median eccentricities of the exoplanets in Fig.~\ref{figP-epla} seem 
approximately constant.
A test confirms this impression~: the Spearman coefficient of the 52 planets
with
$P > 20$~days and $e<0.78$ is 0.21, providing a threshold between 10 and
20~\%. It is thus not possible to rule out the hypothesis that, apart from
circularization due to tidal effects,
the distribution of the eccentricities of the planets is the same for any
period between 20 and 2200 days. Therefore, if the
eccentricities are modified by migration, they 
are changed almost independently of the periods.
However, this applies essentially to periods longer than 200 days,
since we have very few planets between 20 and 200 days.

\subsection{($P$ -- $e$) diagram and metallicities}

Santos et al. (\cite{Santos04}) have shown that
the percentage of stars harboring planets jumps from less than 5~\%
to more than 20~\% when stars with [Fe/H] larger than 0.2 are considered.
In order to see if this limit of 0.2 dex also corresponds to other orbital
properties, we use the [Fe/H] of Santos et al. to distinguish the planets
orbiting ``metallic'' stars and the others. In the ($P$ -- $e$) diagram, we
count 27 ``metallic'' planets, of which 14 have eccentricities smaller
than the median. We conclude with a significance of 79~\%
that metallicity is not related to the eccentricity, confirming the result
obtained by Santos et al. (\cite{Santos03}) with another test.

\section{The planets compared to the binaries}
\label{comparison}

\subsection{The period--eccentricity diagram}
\label{SecPediag}

\begin{figure}
\epsfxsize=8. cm \epsfbox[25 95 455 410]{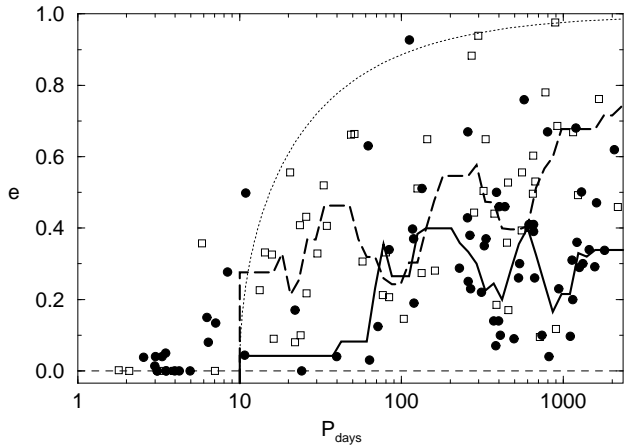}
  \caption[]{The exoplanets compared to the 
non-twin binaries in the ($P$ -- $e$)
diagram. The circles refer to the planets, and the open squares to 
the SB with $q \leq 0.8$. The thin dotted line
is $e_{\rm Max}$, the maximum eccentricity when the circularization
limit is $P_{\rm cutoff}=10$~days. The median
eccentricities are derived after discarding the systems above this limit, in
order to have the same selection effects for both samples. The median
eccentricity of the planets is given by a full line, and that of the SB by a
a dashed line.}
  \label{figP-eplaSB}
\end{figure}

Non-twin binaries and planets are plotted in the ($P$ -- $e$)
diagram in Fig.~\ref{figP-eplaSB}. The median eccentricities
are derived from the systems with $e<e_{\rm Max}$.
In order to make the comparison
free of differences in the tidal circularization, $e_{\rm Max}$ is derived from
Eq.~(\ref{eMax}) assuming $P_{\rm cutoff}=10$~days for both samples. It appears
clearly
that, although all objects are distributed in the same area of the ($P$ -- $e$)
diagram, the planets have eccentricities that are on average smaller than those
of the SB. The test of the distribution around the common median confirms this
discrepancy~: Among 53 planets included in a sample of 102 objects, we count
33 planets below the median eccentricity. The null hypothesis is
rejected at the 1.7~\% level of significance.

It appears from Fig.~\ref{figP-epla} that several planets with long periods and
minimum masses below 2 Jupiter have very small eccentricities. Therefore,
although we have seen that the ($P$ -- $e$) relation may be the same
for all planets, it is relevant to
compare the SB only to the planets with masses larger than 2 Jupiter.
When the planets above 2 Jupiter masses are compared with all the non-twin
SB, the null hypothesis is rejected again at the
4.8~\% level of significance, confirming that the heavy--mass planets have
less eccentric orbits than the binaries.

\subsection{The intrinsic distribution of eccentricities}
\label{secfe}

A direct comparison of the distributions of the eccentricities of
planets and of SB is not feasible, since the possible range of
eccentricities varies with the period, and the distribution of $P$ is not
the same for planets as for SB. Fortunately, another approach may be used
to visually compare these objects, which is to derive the original
distributions of eccentricities corrected for the bias coming from tidal
circularization.
The low significance values of the Spearman correlation tests performed above
allows us to assume that, apart from the area affected by circularization,
the eccentricity distribution does not vary with the period. Therefore, it
is possible to derive the intrinsic distribution of the eccentricities for
planets and for SB, using the method of the ``nested boxes''.
For that purpose, we use Eq.~(\ref{eMax}) to compute, for each system,
the maximum eccentricity unaffected by tides, $e_{\rm Max}$.
We then apply the method given in the Appendix. 
The results are shown in Fig.~\ref{fige}.

\begin{figure}
\epsfxsize=8. cm \epsfbox[25 95 455 410]{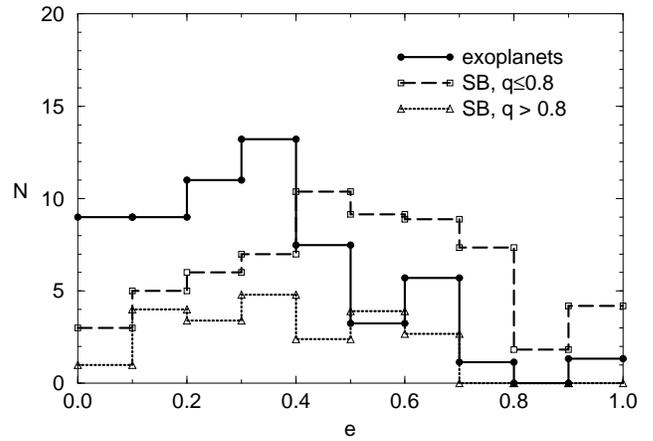}
  \caption[]{Distribution of the eccentricities of SB and planets corrected
for tidal circularization.}
  \label{fige}
\end{figure}

The frequency of planets with low eccentricities is mainly due to the planets
with masses below 2 Jupiter. The largest difference between the two groups
of planets is obtained for $e=0.20$ exactly~: 40~\% of the planets below 2
Jupiter masses have orbits with
$e \le 0.20$, instead of 18~\% for the others.
However, a Smirnov test indicates that this discrepancy is far from
sufficient to reject the hypothesis that all planets obey the same
distribution, since the rejection threshold is as large as 38~\%.
An anonymous referee wondered whether this test could be affected by a bias
related to the radial velocity semi-amplitude. However, this bias is
unfavorable to the planets with the lowest masses, since they are the
most difficult to detect, and its efficiency is maximum when they have
eccentric orbits (see Sect.~\ref{simubias} hereafter). As a consequence,
correcting this bias would slightly decrease
the proportion of $e \leq 0.2$ among the planets below 2 Jupiter, and
then still increase the threshold of the test.
We conclude again that we can assume that the eccentricity distribution of the
planets does not depend on mass.

The differences pointed out by the tests and by the comparison of the medians
are clearly visible in this figure. The maximum of the distribution is
around 0.4 or 0.5 for the non-twin SB, and between 0.3 and 0.4 for the
planets. Moreover, the distribution of $e$ decreases rapidly after the
maximum for the planets, but it is rather flat until 0.8 for the non-twins SB.
A comparison between the median $e$ of the exoplanets and of the twins
could suggest that these two kinds of objects have similar distributions of
eccentricities. However, differences appear in Fig.~\ref{fige}~:
instead of exhibiting a maximum like the exoplanets,
the distribution of eccentricities of twins is nearly flat over a wide
range, from $e=0.1$ to $e=0.7$.

\subsection{Is this difference real ?}
\label{simubias}

We now want to check that the lack of planets with large eccentricities is
not due to a selection effect against the detection of these systems.
In contrast to the SB, 
the detection of planets is far from complete, and our sample is
obviously biased in favor of those which are easiest to detect. The most
obvious bias
is against the detection of long period systems, but this does not affect
the reliability of our test based on the median $e$;
it just decreases the contribution of the long-period planets.
However, another bias is directly related to
the eccentricity~: a large eccentricity increases the semi-amplitude in RV,
but, at the same time, it decreases the $rms$ of the RV measurements.
Therefore, the detection of a system close to the limit of the
instrument is more difficult when the eccentricity is large.
Another consequence of this effect is a bias in the distribution of the
periastron longitude, $\omega$. The orbits with $\omega$ around 0 or $\pi$
are more difficult to detect than those with $\omega$ around $\pi/2$ or
$3\pi/2$. This is visible, although not very significant, in our sample of
planets: we count only 19
orbits with $\omega \in [-\pi/4,+\pi/4] \cup [3\pi/4, 5\pi/4]$
among 43 planets with $e_{\rm Max}>e>0.1$ (the orbits with $e<0.1$ are
not taken into account since $\omega$ is then not reliable).

Simulations have been performed to investigate if this bias
may explain the discrepancy
between planets and SB. Each planet receives the eccentricity
of a SB, randomly taken among the 10 SB with periods closest to that of
the planet. The periastron longitude of the planet is randomly generated, and
3 radial velocity measurements are produced for 3 epochs randomly chosen,
adding errors drawn from the residual $rms$ of the true orbit;
(in reality, each star observed for planet detection receives much more
than 3 observations, but our aim is to derive an upper limit to the bias).
When the standard deviation of
the simulated RV is larger than the threshold corresponding to
$P(\chi^2)=1$~\%, the planet is counted as detected; if not, another
eccentricity is generated, and the simulation of the detection is performed
again, until the planet satisfies the detection condition. When the
complete sample has thus been detected by the simulation, the test of the median
eccentricity in the ($P$ -- $e$) diagram is performed. The simulation of the
diagram is repeated 50\,000 times.

It appears from this calculation that the effect of the bias is to
shift on average 0.6 more planets below the common median.
Assuming that the number of planets below the median would be 32 in the
absence of bias (instead of 33, see Sect.~\ref{SecPediag}), the rejection
threshold of $H_0$ becomes 4.7~\%.
This is still small enough to maintain rejection. At the same time, we count
the planets with $\omega \in [-\pi/4,+\pi/4] \cup [3\pi/4, 5\pi/4]$ which
are detected in the simulation. Among the planets with $e>0.1$, their
proportion is 45.5~\%, in very nice agreement with the observed one, which
is 19/43 = 44~\%. We conclude then that the bias against detection of
orbits with large eccentricities cannot explain the excess of planets with
$e$ smaller than the common median in the ($P$ -- $e$) diagram.

\section{Discussion and conclusion}
\label{conclusion}

We have found some relevant features in comparing the eccentricities of the
SB to those of the exoplanets~:

\begin{itemize}
\item
The ($P_{\rm sr}$ -- $e$) diagram, based on $P_{\rm sr}$ defined in Eq.~(\ref{Pcirc}),
is a powerful tool for determining the limit between the
circularized orbits and the others. The contrast between the two areas in the
diagram suggests that the tidal effects are efficient only when the semilatus
rectum of the orbit is less than a fixed limit. 

\item
The transition from the circularization to the part not affected by tidal
effects looks sharper for the exoplanets than for the SB. For the SB,
it corresponds to  $P_{\rm sr}$ between around 5 and 10 days, in
agreement with Mathieu \& Mazeh (\cite{MaMa88}), Duquennoy et al.
(\cite{DMM92}),
Mathieu et al. (\cite{MaDuLaM92}) and Witte \& Savonije (\cite{WiSa02}),
who consider
that circularization is not restricted to the time of binary formation, but
is still progressing during the whole lifetime of the main sequence components.
For the exoplanets, the fast transition observed for planets with
different ages is consistent with the idea that the tides were efficient
only during the
formation of the system. The inefficiency of tides for a formed planet is in
agreement with the circularization time
derived by Zahn (\cite{Zahn77}), which is a function of $(1+q)/q$.
If the planets were
brought closer to their host stars by migration, that means that migration
occurred when the formation of the planets or of the host stars was not
completed. Tidal circularization was then dominated by the tidal bulge on the
planet, which was hotter, and therefore larger than it is today. 

\item
Before tidal effects had modified them, the eccentricities of planets or
binaries were not strongly related to the periods, or to the angular
momenta. It is even quite possible that they were not correlated at all
with these parameters, since the absence of correlation is not clearly
rejected by statistical tests, and also because our assumption that
tides did not affect at all the orbits with $P_{\rm sr} > P_{\rm cutoff}$ is
possibly a bit too simple. Since the planets in the sample are supposed to have
migrated, this suggests that migration did not alter the
eccentricities significantly; alternatively, it is possible that the
eccentricities were modified, but almost independently of the final periods.

\item
The exoplanets have orbits with eccentricities significantly smaller than
those of the SB 
with the same period and with mass ratios larger than 0.8 (the non-twin
binaries). A similar feature has already been
pointed out for $P<50$~days (Udry \cite{Udry01}), but neglecting
the difference between the distributions of periods of binaries and of planets.
Moreover, it is now certain that the low eccentricities of planets are not 
an effect of the selection of the observed sample. Additionally, 
it seems that the distributions of the eccentricities are not related
to the masses of the companions, neither among the non-twin binaries, 
nor among the planets. Therefore, this discrepancy is probably not
an effect of the low masses of planets in
a formation/evolution process common to planets and binaries~:
this would imply a process depending on the secondary mass,
but only around the transition between stellar and planetary
companions. In fact, the SB most
similar to the planets are the twins, perhaps because these systems were also
interacting with a disk at the time of their formation, as
Tokovinin (\cite{Toko00}) suggested.

\end{itemize}

Our most relevant conclusion is that the eccentricities of
the exoplanet orbits are rather in favor of the hypothesis
that exoplanets and binary stars are
not the products of the same physical process. After the 
``brown dwarf desert'' (Halbwachs et al. \cite{Halb00}) separating the stellar
components from the
planets in the distribution of the secondary masses, it is another
argument in that sense which was derived from statistical investigations.

\begin{acknowledgements}
We are grateful to Piet Hut and Jean-Paul Zahn for their explanations
and for their valuable comments on the draft version of the paper.
Douglas Heggie carefully read a preliminary version and added relevant
corrections. 
An anonymous referee made valuable comments. The A\&A language editor, Jet
Katgert, corrected the English. The selection of the exoplanets was partly
based on data taken from
Simbad, the database of the Centre de Donn\'ees astronomiques de Strasbourg.
\end{acknowledgements}

\appendix

\section{The method of the nested boxes applied to the distribution of
eccentricities}

This method
was initially dedicated to the derivation of a
bias-free distribution of mass ratios of visual binaries
(Halbwachs \cite{Halb83}, Halbwachs et al. \cite{Halb97}), as is 
explained in detail in Halbwachs (\cite{Halb01}). It is adapted hereafter
to the derivation of the intrinsic distribution of the eccentricities.

We consider a sample with periods $P$ larger than $P_{\rm cutoff}$, the period
corresponding to tidal circularization, as explained in Sect.~\ref{tides}. 
For each system, $P$ and $P_{\rm cutoff}$ are used to derive $e_{\rm Max}$, the maximum
eccentricity of the
orbits unaffected by tidal effects. The systems having eccentricities 
$e > e_{\rm Max}$, if any, are discarded from the sample.

The principle of method is as follows~: 
\begin{itemize}
\item
We define a first ``box'' by setting
a small minimum value of $e_{\rm Max}$, called $e_1$ hereafter. The eccentricities
of the systems having $e_{\rm Max} \geq e_1$ are used to derive a first estimation
of the intrinsic distribution 
of eccentricities, $f_1(e)$, which is defined from $e=0$
up to $e=e_1$. Since $e_1$ is small, this distribution is based on a large
number of systems, and it is fairly reliable; unfortunately, it concerns only
a small range of eccentricities. 
\item
A second box is defined, using a limit
$e_2 > e_1$. The eccentricities of the systems with $e_{\rm Max} \geq e_2$ provide
a second estimation of the distribution, $f_2(e)$, which applies to $e \le e_2$.
We have then an estimation of $f(e)$ which is valid between $e_1$ 
and $e_2$, but for $e \leq e_1$, $f_2(e)$ is less reliable than $f_1(e)$, since
it is derived from fewer systems in this range. 
\item
The two boxes defined above are ``nested'', since the systems with $e \leq e_1$
belonging to the second box are all also present in the first box. This common
part is used to derive $f(e)$ by connecting $f_2(e > e_1)$ to $f_1(e \leq e1)$.
Let $N_1$ and $N_2$ be the numbers of systems in the first box and in the second box,
respectively. If the second box contains $n_2(e_1)$ systems with $e \leq e_1$, the
best estimation of $f(e)$ is~:

\begin{equation}
f_{1,2} (e) = f_1(e) + \theta(e-e_1) \frac{N_1}{n_2(e_1)} f_2(e)
\label{nest1}
\end{equation}

\noindent
where $\theta$ is the Heaviside function; it is unnecessary to normalized the
distributions so early in the 
calculation. 

\item
We may still add several boxes, using limits $e_i$ with increasing values. If
$N_{i-1}$ is the norm of the distribution derived from the boxes $1$ to $i-1$, 
$f_{1, i-1}$, and if box $i$
contains $n_i(e_{i-1})$ systems with $e \leq e_{i-1}$, Eq.~(\ref{nest1})
becomes~:

\begin{equation}
f_{1, i} (e) = f_{1, i-1}(e) + \theta(e-e_{i-1}) \frac{N_{i-1}}{n_i(e_{i-1})} f_i(e)
\label{nesti}
\end{equation}

The distribution is normalized after adding the last box.

\end{itemize}

In practice, the values of the $e_i$ terms are $i \times 0.01$ when the
eccentricities are provided with two decimals, in order to take into
account all the systems with $e \leq e_{\rm Max}$. At the end of the calculation,
the 0.01--bins are merged into 0.1--bins, in order to
make the final distribution more readable.

\end{document}